\documentclass[prl,twocolumn,preprintnumbers,superscriptaddress]{revtex4-2}
\usepackage{amssymb}
\usepackage{amsmath}
\usepackage{amsfonts}
\usepackage{wasysym}
\usepackage{graphicx}
\usepackage{dcolumn}
\usepackage{bm}
\usepackage{color}
\usepackage[colorlinks,linkcolor=blue, urlcolor=blue, anchorcolor=blue, citecolor=blue]{hyperref}

\def\GJ{\textcolor{black}}

\begin{document}
	
	\title{Coherent Control of Collective Spontaneous Emission through Self-interference}
	
	\author{Lei Qiao}
	\email{qiaolei@nus.edu.sg}
	\affiliation{Department of Physics, National University of Singapore, Singapore 117551, Singapore}
	
	\author{Jiangbin Gong}
	\email{phygj@nus.edu.sg}
	\affiliation{Department of Physics, National University of Singapore, Singapore 117551, Singapore}

\begin{abstract}
As one of the central topics in quantum optics, collective spontaneous
emission such as superradiance has been realized in a variety of systems.
This work proposes an innovative scheme to coherently control collective
emission rates via a self-interference mechanism in a nonlinear waveguide
setting. The self-interference is made possible by photon backward
scattering incurred by quantum scatterers in a waveguide working as quantum
switches. Whether the interference is constructive or destructive is found
to depend strongly on the distance between the scatterers and the emitters.
The interference between two propagation pathways of the same photon leads
to controllable superradiance and subradiance, with their collective decay
rates much enhanced or suppressed (also leading to hyperradiance or
population trapping). Furthermore, the self-interference mechanism is
manifested by an abrupt change in the emission rates in real time. An
experimental setup based on superconducting transmission line resonators and
transmon qubits is further proposed to realize controllable collective emission rates. 
\end{abstract}

\maketitle

\emph{Introduction.}---Waveguide quantum electrodynamics has recently been a
growing area in quantum optics with important applications in quantum
information processing \cite{Gu17,Wendin17,Roy17,Shen05,Zhou08}. Different
integrations of quantum emitters (QEs) with nanophotonic structures have
been achieved, such as guided surface plasmons confined on a conducting
nanowire with individual optical emitters \cite{Chang07,Rycenga11}, photonic
nanowire coupled by embedded quantum dots \cite{Claudon10,Lodahl15}, or
superconducting transmission line with superconducting qubits \cite%
{Astafiev10,You11,Houck12}. These physical platforms make it possible to let
QEs interact with one-dimensional bosonic modes with nontrivial dispersions,
leading to intriguing dynamics such as persistent quantum beats \cite%
{Zheng13,Song21}, supercorrelated radiance \cite{Wang20}, and single photons
by quenching the vacuum \cite{Burillo19}. Indeed, enhanced light-matter
interaction because of dimensionality reduction in waveguide quantum
electrodynamics continues to attract a great deal of attention \cite%
{Guo20,Andersson19,Pichler16,Facchi18,Tudela17,Garmon19,Calajo19,Dinc19,Burillo17,Qiao19A,Qiao20}%
. Advances in designing and probing light-matter interactions have allowed
the investigation of collective phenomena such as superradiance, subradiance 
\cite%
{Dicke54,Gross82,Sinha20,Svidzinsky08,Slepyan13,Jenkins17,Ke19,Zhang19,Bienaime12}
, cavity antiresonance spectroscopy \cite{Plankensteiner17}, and
nonequilibrium collective phase transition \cite{Bakhtiari2015,Yamamoto21}.

As a typical collective emission, Dicke superradiance has been demonstrated
in systems of hot atoms \cite{Skribanowitz73,Gross73}, cold atoms \cite%
{Inouye99,Baumann10}, trapped ions \cite{DeVoe96,Begley16,Casabone15}, and
superconducting qubits \cite{Mlynek14,Filipp11}, etc. Its counterpart with
reduced emission rate is termed subradiance. Transitions between
superradiant and subradiant states have been realized in superconducting
circuits by initially applying phase gate on each qubit \cite{Wang20exp}.
Superradiance and subradiance are highly relevant to quantum memory as their
roles can be important in the writing and reading of quantum information 
\cite{Mendes13,Scully15}. However, to date continuously controllable
collective emission rate without using external driving fields \cite{Woldeyo19,Woldeyo03,Jiang11}
 remains a challenge for almost all QE systems.

In this letter we reveal an unknown aspect of spontaneous emission in a
nonlinear waveguide setting. We consider quantum scatterers in addition to
general quantum emitters in the same waveguide. The emitted photon
propagating in the waveguide can be bounced back by the scatterers as
quantum switches. The backward scattered photon then interferes with the
other branch of the photon propagating in the opposite direction. Such
self-interference is exploited to achieve continuous and extensive control
of the spontaneous emission rate of QEs. As shown below, even the
transition from superradiance to subradiance can be readily achieved if we
control certain features of the quantum switches, such as its resonance
frequency and the QE-scatterer distance. The underlying self-interference
mechanism is analyzed both qualitatively and quantitatively, with excellent
agreement between physical insights and theoretical studies. In particular,
the QE-scatterer distance is found to be a
crucial parameter to induce constructive or destructive interference. An
experimental setup based on superconducting transmission line resonators and
transmon qubits is further proposed to realize continuously controllable collective emission rates. 

\begin{figure}[t]
\begin{center}
\includegraphics[width=\columnwidth]{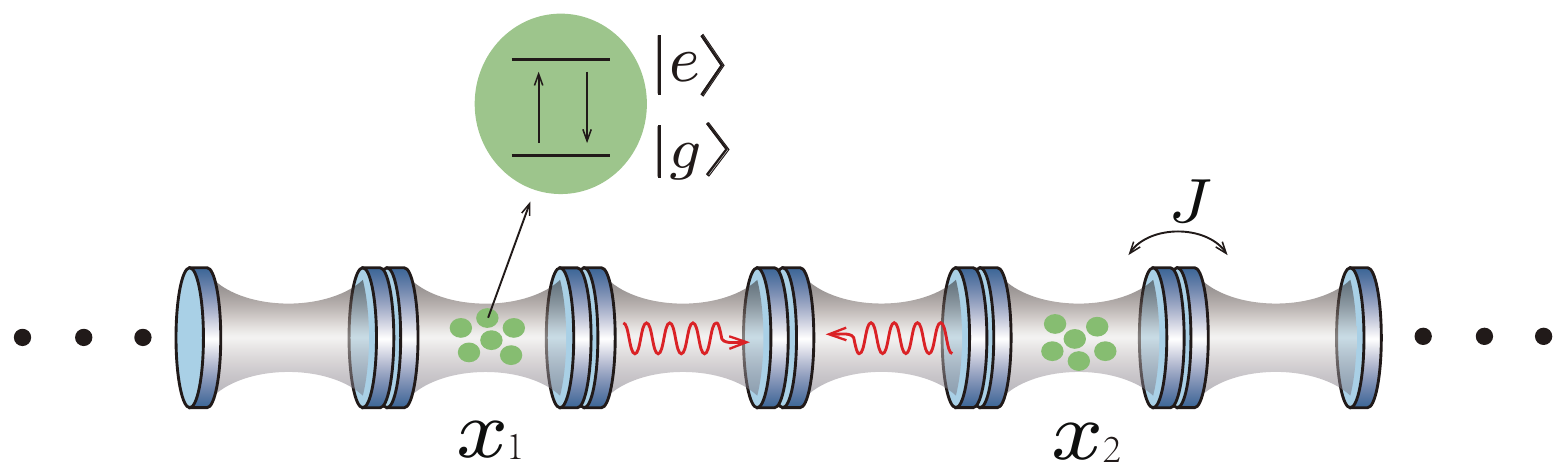}
\end{center}
\caption{Schematic of a waveguide setup. A one-dimensional array of
resonators with nearest-neighbor tunneling $J$ hosts an assembly of quantum
emitters at $x_{1}$ and a second collection of two-level atoms as quantum
scatterers at $x_{2}$.}
\label{fig1_Setup}
\end{figure}

\emph{Model.}---The system we consider consists of a one-dimensional array
of tunneling-coupled cavities which accomodate one assembly of QEs at
position $x=x_{1}$, as well as a second collection of two-level atoms at $%
x_{2}$ respectively. A schematic plot of this configuration is shown
in Fig.~\ref{fig1_Setup}. Atoms at $x_{2}$ 
play the role of quantum scatterers, through which the spontaneous emission
dynamics of QEs at $x_{1}$ is to be manipulated. Though playing two
different roles, these two collections of atoms will be treated with similar
notation, indexed by $A$ and $B$ respectively, and assumed to have excited
states $|e^{A}\rangle $, $|e^{B}\rangle $ and ground states $|g^{A}\rangle $%
, $|g^{B}\rangle $, separated in energy by frequencies $\Omega _{A}$ and $%
\Omega _{B}$ (we set $\hbar =1$ throughout). The tunneling-coupled photonic
waveguide forms a lattice, modeled by the following tight-binding Hamiltonian%
\begin{equation}
H_{\text{ph}}=\sum_{x}\omega _{c}a_{x}^{\dag }a_{x}+J\sum_{x}\left(
a_{x+1}^{\dag }a_{x}+a_{x}^{\dag }a_{x+1}\right) \text{,}  \label{Hph}
\end{equation}%
where $a_{x}^{\dag }$ is the creation operator of the waveguide mode at
position $x$, and $\omega _{c}$ is the resonance frequency of a single
cavity. For convenience, we assume the lattice constant to be $a=1$
throughout. The total Hamiltonian describing the system is then%
\begin{align}
H& =H_{\text{ph}}+\sum_{j}\Omega _{j,A}\left\vert e_{j}^{A}\right\rangle
\left\langle e_{j}^{A}\right\vert +\sum_{j}\Omega _{j,B}\left\vert
e_{j}^{B}\right\rangle \left\langle e_{j}^{B}\right\vert  \notag \\
& +\sum_{j}\left( V_{j,A}\sigma _{j,A}^{+}a_{x_{1}}+V_{j,B}\sigma
_{j,B}^{+}a_{x_{2}}+h.c\right) \text{,}  \label{totalH}
\end{align}%
where $\sigma _{j,A}^{+}$ and $\sigma _{j,B}^{+}$ are the creation operators
for the $j$th atom in each assembly and $V_{j,A}$ and $V_{j,B}$ are the
respective coupling strengths. For the case of only one QE with weak
coupling ($V\ll J$), the QE decays with the radiation rate $\Gamma =V^{2}/J$
if the emitter is near resonance with the frequency of a single cavity \cite%
{Lombardo14}. In the single excitation subspace, the spectrum of Eq. (\ref%
{totalH}) comprises discrete localized bound states and a continuum of
delocalized dressed states with energy $\omega _{k}=\omega _{c}+2J\cos
\left( k\right) $ vs the mode wavevector $k$, thus forming a scattering band
with $\omega _{c}$ being the band central frequency with bandwidth $%
4J$ ($J>0$). The said bound states result in the known fractional trapping of
an emitted photon and nonexponential dynamics of the spontaneous emission 
\cite{John94,Qiao19}. If the \GJ{excitation} frequency of QEs is equal to the single-cavity
frequency $\omega _{c}$ (a condition assumed below), then the momentum of
a \GJ{radiated} photon is around $k_{m}=\pm \pi /2$. Note also that the peak photon
group velocity is $|v_{g}^{m}|=2J$ reached by the wavevector $k_{m}$.

\emph{General theoretical considerations.}---Let us now assume that the QEs
and the quantum scatterers are separated by a distance $\Delta x\equiv
|x_{2}-x_{1}|$, which is less than the half of the coherence length $\sim
v_{g}^{m}/\Gamma $ of a spontaneously emitted photon. The initial state of
the whole system is that one of the QEs is excited, or a superpositions of
such configurations, with the photon field in vacuum. This hence places the
whole wavefunction in the single-excitation invariant subspace. The
time-evolving state at time $t$ can be written as%
\begin{equation}
\left\vert \psi \left( t\right) \right\rangle =\left\{
\sum_{i}\sum_{j=1}^{M_{i}}C_{j}^{i}\left( t\right) \sigma
_{j,i}^{+}+\sum_{k}C_{k}\left( t\right) a_{k}^{\dag }\right\} \left\vert g%
\text{,vac}\right\rangle \text{,}  \label{OriDynFunc}
\end{equation}%
where $i=A,B$ and $a_{k}^{\dag }=(1/\sqrt{N})\sum_{x}e^{ikx}a_{x}^{\dag }$. $%
M_{i}$ is the number of the QEs or of the scatterers. $C_{j}^{i}$ is the
excitation amplitude for the $j$th atom in each collection of atoms, $C_{k}$
is the amplitude of the waveguide mode with momentum $k$. Without loss of
generality, we assume that only QEs with indices $j_{n}$ may be excited at
time zero, i.e., $C_{j_{n}}^{A}(0)\neq 0$, $C_{j}^{A}(0)=0$ ($j\neq j_{n}$),
and $C_{j}^{B}(0)=C_{k}(0)=0$. If initially at most two excited QEs indexed
by $j_{1}$ and $j_{2}$ are involved in the initial excitation, then
exact results about their ensuing time dependence can be obtained:  \cite{Supple}

\begin{align}
C_{j_{1}}^{A}\left( t\right) & =\frac{L_{1}(s)C_{j_{1}}^{A}\left( 0\right)
+L_{2}(s)C_{j_{2}}^{A}\left( 0\right) }{G(s)}e^{st}|_{s=-i\Omega _{A}} 
\notag \\
& +\sum_{m}\frac{iL_{2}(s)\left[ C_{j_{1}}^{A}\left( 0\right)
+C_{j_{2}}^{A}\left( 0\right) \right] }{\left( is-\Omega _{A}\right)
[G(s)]^{\prime }}e^{st}|_{s=x_{m}}  \notag \\
& -\sum_{\alpha =\pm }\int_{-1}^{1}\frac{\left[ Q_{1}^{\alpha
}C_{j_{1}}^{A}\left( 0\right) +Q_{2}^{\alpha }C_{j_{2}}^{A}\left( 0\right) %
\right] e^{i2Jyt}}{2\pi i\alpha \left( y+\Omega _{A}\right) [Q_{1}^{\alpha
}-Q_{2}^{\alpha }]}dy\text{,}  \label{Result1}
\end{align}%
where $L_{1,2}(s)$ and $G(s)$ strongly depend on the separation parameter $%
\Delta x$, $Q_{1,2}^{\pm }$ are explicit functions of the integration
variable $y$, and $x_{m}$ is the roots of the equation $G(s)=0$ \cite{Supple}%
. The imaginary part of each $x_{m}$ corresponds to the inverse of system's
eigenenergies of the localized photon-QE dressed states. In fact, the second
term on the right-hand side of Eq. (\ref{Result1}) originates from the
system's photon-QE bound states with nonzero field amplitudes. $C_{j_{2}}^{A}(t)$ can be obtained by exchanging the positions
of $C_{j_{1}}^{A}(0)$ and $C_{j_{2}}^{A}(0)$ in Eq.~(\ref{Result1}). In
obtaining the analytical expressions above, it has been assumed that each
collection of atoms (namely, among the QEs, or among the scatterers) are
identical and thus $\Omega _{j,i}$ and $V_{j,i}$ are independent of the atom
index $j$, denoted by $\Omega _{A}$, $\Omega _{B}$, $V_{A}$, and $V_{B}$.

\begin{figure}[t]
\begin{center}
\includegraphics[width=\columnwidth]{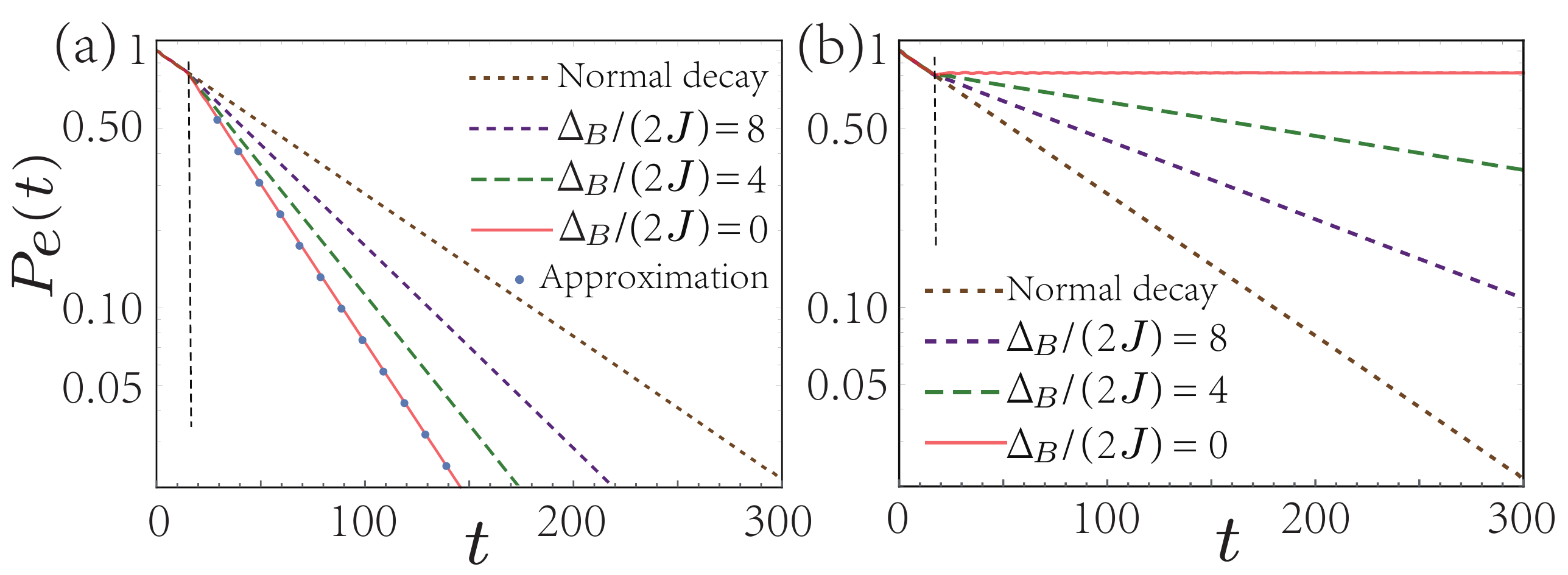}
\end{center}
\caption{(a) Excited state population $P_{e}(t)$ vs  detuning $\Delta _{B}$ for $\Delta x=7$. (b) Excited state population  $P_{e}(t)$ vs detuning $\Delta
_{B} $ for $\Delta x=8$.  Time is in units of $1/(2J)$. The parameters
are $V_{A}/(2J)=0.08$, $V_{B}/(2J)=1.8$, $\Delta _{A}/(2J)=0$, the number of
emitters is $M_{A}=1$ and that of scatterers is $M_{B}=2$. In both (a)
and (b) the vertical dashed lines indicate the time of arrival of a
reflected photon. }
\label{fig2_radiance}
\end{figure}

\emph{Cases with one single emitter.}---To gain some important insights
first, we consider a single QE indexed by $j_{1}$ coupled with mutiple
scatterers through the waveguide, with the initial state $|\psi (0)\rangle $ 
$=$ $\sigma _{A}^{+}|g$,vac$\rangle $. The time evolution of the excited
state population $P_{e}(t)=|C_{j_{1}}^{A}\left( t\right) |^{2}$ is shown in
Fig.~\ref{fig2_radiance}(a) vs the detuning parameter $\Delta _{B}=\Omega
_{B}-\omega _{c}$ depicting the scatterers,  with the QE-scatterer separation $\Delta x=7$ (i.e., 7
lattice constants) as an example. At early times, the emission dynamics
matches well with that of a normal decay process $P_{e}(t)\approx e^{-\Gamma
_{1}t}$ with $\Gamma _{1}=V_{A}^{2}/J$ (assuming $V_{A}\ll J)$, without
feeling the presence of the scatterers. Later  the
scatterers makes a dramatic difference during the spontaneous
emission process. In particular, as the main reason to introduce the
scatterers in the first place, the scatterers as two-level systems can
extensively control the coherent transport of a single photon in the
waveguide, including a complete reflection of the emitted photon \cite%
{Zhou08}. Hence, once part or even the whole of the propagating photon
towards the scatterers comes back to the decaying QE, it will interfere with
the other branch of the photon propagating along the other direction.
Supporting this physical picture, Fig.~\ref{fig2_radiance}(a) depicts an
abrupt change of the radiation rate once $t$ reaches $t_{0}\approx 2\Delta
x/(2J)$, yielding a cascade of stimulated emission. In terms of the
real-time dynamics, we are witnessing an intriguing scenario that an emitted
photon, when being bounced back, can further boost the emission process that
has not been completed yet. The degree of enhancement is continuously
controlled by the frequency of the scatterers, as evidenced by the shown
strong dependence of the emission rates vs the detuning parameter $\Delta
_{B}$.

Continuing our investigations of the cases with $\Delta x=2n+1$ ($n$ being
an integer), let us further assume that the quantum scatterers are near
resonance with each single cavity with small $\Delta x$, plus the conditions 
$V_{A}\ll 2J$ and $\sqrt{M_{B}}V_{B}\sim 2J$. The enhanced emission rate
under these conditions are found to be $P_{e}(t)\approx \lbrack
M_{B}V_{B}^{2}/(4J^{2})-\alpha _{1}/2]^{2}e^{-2J\alpha _{1}t}/\beta _{1}$
with $\alpha _{1}=V_{A}^{2}/(J^{2}-\Delta xV_{A}^{2}/2)$ and $\beta _{1}$ is
a time-independent quantity \cite{Supple}. As can be seen in Fig.~\ref%
{fig2_radiance}(a), our approximate theoretical results agree well with the
exact results obtained from Eq.~(\ref{Result1}). For $\Delta x$ being small
enough, the radiance rate is two times the normal decay rate. Although there
is only one QE here, the emission rate is much enhanced and even beyond
two-QE Dicke superradiance. The physical understanding is the following: The
QE interferes with itself via a delayed photon, thus it can effectively
realize collective interference and hence achieve superradiance.

\begin{figure}[t]
\begin{center}
\includegraphics[width=\columnwidth]{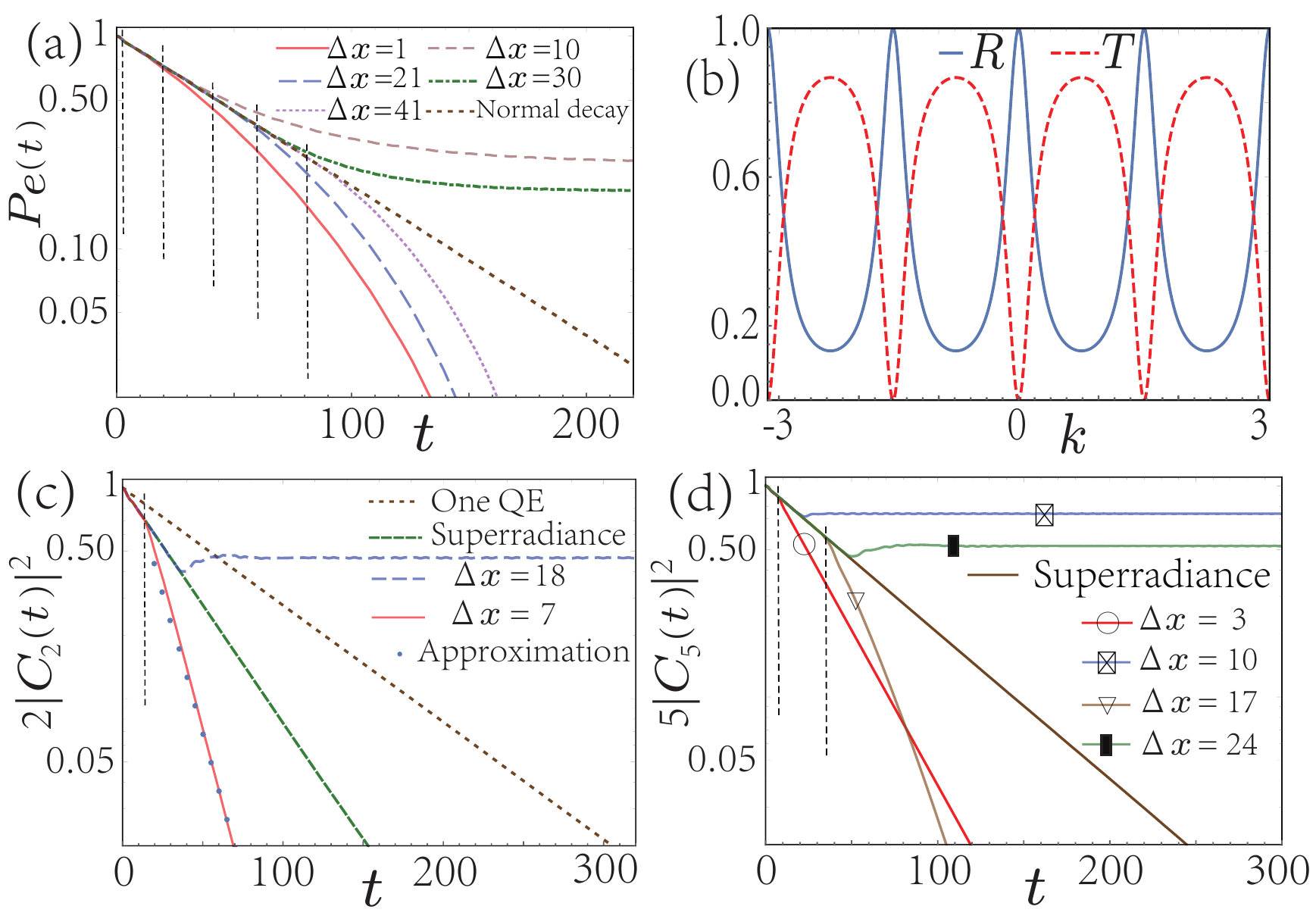}
\end{center}
\caption{(a) Excited state population $P_{e}(t)$ vs $\Delta x$, with $V_{A}/(2J)=0.09$, $%
V_{B}/(2J)=0.07$, $\Delta _{A}/(2J)=0$, $\Delta _{B}/(2J)=0$, $M_{A}=1$ and $%
M_{B}=2$. (b) Reflection coefficient $R=\left\vert r_{k}\right\vert ^{2}$
and transmission coefficient $T=|1+r_{k}|^{2}$ as a function of momentum $k$%
. Parameters are $\Delta _{B}/(\protect\sqrt{M_{B}} V_{B})=0$, $J/(\protect%
\sqrt{M_{B}}V_{B})=1.13$. (c) Excited state population 
$2|C_{\text{2}}\left( t\right) |^{2}$, with $\Delta
_{A}/(2J)=0 $, $\Delta _{B}/(2J)=0$, $V_{A}/(2J)=0.08$ , $V_{B}/(2J)=1.27$, $%
M_{A}=2$ and $M_{B}=2$. (d) Excited state population $%
5|C_{\text{5}}\left( t\right) |^{2}$, with $\Delta _{A}/(2J)=0$%
, $\Delta _{B}/(2J)=0$, $V_{A}/(2J)=0.04$ , $V_{B}/(2J)=1.0$, $M_{A}=5$ and $%
M_{B}=2$. Time is in units of $1/(2J)$. Different vertical dashed lines
indicate the different times of arrival of a reflected photon.}
\label{fig3_MoreRadi}
\end{figure}

Next we consider cases with $\Delta x =2n$. A few computational examples with
 $\Delta x=8$ are shown in Fig.~\ref%
{fig2_radiance}(b), where suppressed emission rates are clearly observed. As
the frequency of the scatterers is tuned from $\omega _{c}$ to the values
far away from the photon band, the suppression becomes weak and ultimately
the emission comes back to the normal decay. It is curious to qualitatively
understand why superradiance and subradiance are observed for $\Delta =2n+1$
and $\Delta =2n$ in Fig.~\ref{fig2_radiance}(a) and Fig.~\ref{fig2_radiance}%
(b), respectively.  If $\Delta x=2n+1$,
the phase difference incurred by the round travel of the bounced photon can
be estimated as $|k_{m}|(2\Delta x)=(2n+1)\pi $, if considering the main
wave component around $k_{m}=\pm \pi /2$ with the largest group velocity.
Also accounting for the $\pi $ shift associated with a complete photon
reflection, the overall phase difference between the bounced photon and the
original photon is thus $2n\pi $, yielding constructive interference
and hence enhanced emission. By contrast, if $\Delta x=2n$ is chosen, then
the overall phase difference between the two interfering pathways is $%
(2n+1)\pi $, thus producing destructive interference and leading to
suppression of the emission and thus subradiance. Confirming this
understanding, in Fig.~\ref{fig3_MoreRadi}(a), we further show how the
emission rates for weak coupling $V_{B}$ is changed over a wide range if the
QE-scatterer distance $\Delta x$ is adjusted.

Our results above have clearly indicated the important role of the backward
scattering in the self-interference mechanism. It is hence useful to examine
some details of the scattering process. When the radiation field reaches  the scatterers, 
part of the field is reflected with the
reflection amplitude $r_{k}$ given by $r_{k}=M_{B}V_{B}^{2}[i2J|\sin
(k)|(\omega _{k}-\Omega _{B})-M_{B}V_{B}^{2}]^{-1}$ \cite{Supple}. Around
the resonance, the reflection spectrum yields the so-called Breit-Wigner
line shape with the spectrum width given by $M_{B}V_{B}^{2}/|J\sin (k_{0})|$%
, where $k_{0}$ is determined by the relation $4J^{2}\sin ^{2}(k)(\omega
_{k}-\Omega _{B})^{2}=M_{B}^{2}V_{B}^{4}$ \cite{Zhou08}. In particular, if
the photon energy $\omega _{k}$ is under resonance with the two-level
scatterers, namely, $\omega _{k}=\Omega _{B}$, one then obtains complete
reflection with $r_{k}=-1$ (hence the above-mentioned $\pi $ phase shift).
Under the parameter setting $\omega _{c}=\Omega _{B}$, the resonance
scattering condition $\omega _{k}=w_{c}+2J\cos (k)=\Omega _{B}$ occurs for $%
k=\pi /2$. From the expression of $r_{k}$, it is also seen that if $k=0$ or $%
k=\pi $, complete reflection happens also. However, this is irrelevant to
our self-interference mechanism because such components have a vanishing
group velocity in the waveguide. In Fig.~\ref{fig3_MoreRadi}(b), we show the
reflection and transmission spectra vs momentum $k$.  
%Only the components
%around the $k=\pi /2$ peak can contribute to the self-interference mechanism
%qualitatively explained above, further justifying why we focus on $k_{m}=\pm
%\pi /2$ components in our earlier qualitative reasoning.

\emph{Cases with two and more emitters.}---We now examine how collective
emission rates with two QEs can be manipulated by exploiting
self-interference. The two emitters are prepared in typical
single-photon entangled states $|\psi ^{\pm }\rangle $ $=$ $(1/\sqrt{2}%
)(\sigma _{j_{1},A}^{+}\pm \sigma _{j_{2},A}^{+})|g$,vac$\rangle $. For
 $|\psi ^{-}\rangle $, the decay of QEs is completely suppressed
since it is a dark state that cannot emit a photon. Thus we focus on the
emission dynamics emanating from $|\psi ^{+}\rangle $. When $V_{B}=0$ and
the frequencies of QEs lie outside the scattering band with $|\Delta _{A}\pm
2J|$ $\gg V_{A}$, the evolution of $|\psi ^{+}\rangle $ is dominated by a
trapping regime due to the presence of bound states \cite{John94}. For the
case $V_{A}\ll 2J$, the decay of the amplitudes $%
C_{j_{1}}^{A}(t)=C_{j_{2}}^{A}(t)\equiv C_{\text{2}}\left( t\right) $ is
basically exponential, with a very slowly changing radiation rate as $\Omega
_{A}$ is tuned from from $\omega _{c}\pm J$ to $\omega _{c}$. This is what
one expects from the Wigner-Weisskopf and Markovian perturbative theories,
which predict $|C_{\text{2}}\left( t\right) |^{2}\approx (1/2)e^{-\Gamma
_{s}(\Delta _{A})t}$, with a decay rate $\Gamma _{s}(\Delta _{A})=4\pi
V_{A}^{2}D(\Delta _{A})$, where $D(\Delta _{A})$ is the density of states
for the photon Hamiltonian $H_{\text{ph}}$. $D(\Delta _{A})$
reaches its extremum under the resonance condition $\Delta _{A}=0$ and $\Gamma
_{s}(0)\equiv \Gamma _{s}=2V_{A}^{2}/J$, which is twice the normal decay
rate shown in Fig.~\ref{fig2_radiance}. Indeed, this is what the standard
superradiance theory predicts.

Consider now what happens if the excitation frequency $\Omega _{A}$ of QEs
is around the middle of the band with $|\Delta _{A}|$ $\ll 2J$ in the
presence of scatterers. The frequency $\Omega _{B}$ of the scatterers
controls the position of transmission valley of the radiated photon while $%
M_{B}$ and $V_{B}$ determines the valley width. As in the case of one QE,
before $t$ reaches $t_{0}$, $|C_{\text{2}}\left( t\right) |^{2}$ behaves as
a superradiant state without sensing the quantum scatterers. Beyond $t_{0}$,
it is found that the superradiance rate can be much enhanced if $\Delta
x=2n+1$ and much suppressed if $\Delta x=2n$. Representative results are
shown in Fig.~\ref{fig3_MoreRadi}(c). It is seen that the self-interference
mechanism works effectively in the case of superradiance. In particular, the
enhanced superradiance by constructive self-interference may be termed
hyperradiance, a terminology also used previously but due to different
physics \cite{Pleinert17}. Under the conditions $V_{A}\ll 2J$ and $\sqrt{%
	M_{B}}V_{B}\sim 2J$, we can make reasonable approximations using our general
theoretical expressions and find the following hyperradiance dynamics with
small $\Delta x$  \cite{Supple}:%
\begin{equation}
	C_{\text{2}}\left( t\right) \approx \frac{M_{B}V_{B}^{2}-J\Gamma _{h}}{\sqrt{%
			2\left( \chi ^{2}-16M_{B}V_{A}^{2}V_{B}^{2}\right) }}e^{-\frac{\Gamma _{h}}{2%
		}t}\text{.}  \label{hypereq}
\end{equation}%
Here the radiance rate $\Gamma _{h}=4JV_{A}^{2}/(J^{2}-\Delta xV_{A}^{2})$
with $\chi =M_{B}V_{B}^{2}-\Delta xM_{B}V_{A}^{2}V_{B}^{2}/J^{2}$. The
dotted line shown in Fig.~\ref{fig3_MoreRadi}(c) is obtained from Eq.~(\ref%
{hypereq}), in excellent agreement with exact results obtained directly from
Eq.~(\ref{Result1}). Under the limit $\Delta x$ being small enough, $\Gamma
_{h}$ is found to be just two times the superradiance rate $\Gamma
_{h}=2\Gamma _{s}$. On the other hand, Eq.~(\ref{hypereq}) indicates that $%
\Gamma _{h}$ becomes large as $\Delta x$ increases. As such, tuning $\Delta x
$ allows us to further boost hyperradiance rates.

Finally, we investigate how the self-interference mechanism works when there
are multiple QEs. In this case, we rely fully on computational studies since
it becomes tedious to find analytical results with more than two
QEs being initially excited. To investigate if the above hyperradiance
dynamics can be extended to cases with multiple QEs, we consider the
following initial amplitudes $C_{j_{1}}^{A}(0)=...=C_{j_{M_{A}}}^{A}(0)=1/%
\sqrt{M_{A}}$. Fig.~\ref{fig3_MoreRadi}(d) depicts the results with $M_{A}=5$
QEs for different values of $\Delta x$. In the absence of scatterers, under
the condition $\sqrt{M_{A}}V_{A}\ll 2J$, the amplitudes $C_{j_{1}}^{A}(t)$ =
... = $C_{j_{M_{A}}}^{A}(t)\equiv C_{M_{A}}(t)$ can be approximately
described by $|C_{M_{A}}\left( t\right) |^{2}\approx (1/M_{A})e^{-\Gamma
_{s}t}$ with $\Gamma _{s}=M_{A}V_{A}^{2}/J$, which is nothing but the Dicke
superradiance.

However, for cases with $\Delta x=2n+1$, Fig. \ref{fig3_MoreRadi}(d) shows that the
self-interference mechanism further boosts the collective emission rates by
a factor of two for relatively small $\Delta x$. Furthermore, as $\Delta x$
increases, the emission rate continues to be enhanced and hence surpasses $%
2\Gamma _{s}$. This echoes with our observation in the case of two emitters.
To understand this intriguing trend due to increasing $\Delta x$, we first
note that the wavevevectors of the photon that will be backward scattered
are spread around $k=\pi /2$ (in the direction towards the scatterers), but only
the component with precisely $k=\pi/2$ of the largest group velocity
can optimally induce the constructive self-interference. If $\Delta x$
increases, the potential phase dispersion among these components along the
the propagation pathway increases. This imposes a more strict selection on the wave
components that can contribute to the self-interference. These
selected wavecomponents also tend to induce the self-interference more
synchronously.

For cases with $\Delta x=2n$, it is clearly observed in Fig. \ref%
{fig3_MoreRadi}(d) that the emission rates are much suppressed due to the
destructive self-interference. For suppressed subradiance, the asymptotic
values of the excited state population is finite at sufficiently long time.
Indeed, under $\Delta_B =0$, the emitted photon is first bounced back by the scatterers. 
Once the scattered photon meets the QEs, destructive self-interference suppresses the collective emission and as such the QEs tend to reflect the emitted photon as well, thus also dynamically trapping the photon between the QEs and the scatterers. \GJ{These results and insights
indicate} the role of bound states in fully explaining the population
trapping on the excited state.  
%the emitted photon is now trapped by the scatterers as well as the QEs the  he asymptotic non-zero population on the excited state can be easily
%explained in terms of the existence of some photon-QE bound states, it is
%intriguing that only the destructive
%self-interference can help the system to populate on the bound states.

%Therefore, for multiple QEs, we still observe enhancement of the collective
%emission rates if $\Delta x=2n+1$ and then suppression of the emission rates
%if $\Delta x=2n$. The degree of the enhancement and suppression are also
%seen to be continously controllable by adjusting the frequency of the
%scatterers. Thus, the possible range of the collective emission is widely
%controllable by adjusting the separation $\Delta x$ and the frequency $%
%\Omega _{B}$.

\emph{Discussion and conclusions.}---In a one-dimensional nonlinear
waveguide setting, we have shown that the self-interference incurred by a
backward scattered photon originally emitted from quantum emitters can
dramatically change the emission rate. The backward scattering is caused by
a collection of quantum scatterers with a preselected distance. For an
initial state as a superposition of excitation of not more than two
emitters, we obtain an exact analytical solution to predict how quantum
scatterers can be used to control the emission rate. In both our theoretical
treatments and our qualitative analysis, it is found that the distance
between the emitters and the scatterers plays a critical role in deciding
whether constructive or destructive interference occurs. This
self-interference mechanism then leads to extensive control of the
collective emission dynamics, ranging from hyperrardiance to strongly
suppressed subradiance. The theory and the physics communicated in this work
are rather general. Indeed, the considered quantum emitters can be of
different types, including both natural and artificial ones.

It might not be straightforward to tune the separation between the emitters
and the scatterers if they are already grown in a nanophotonic structure.
We propose to install several different
groups of quantum scatterers at different positions in the waveguide. When
the transition frequencies of the scatterers are tuned to be outside the
band and far away from the resonance frequency of a single cavity, this
group of scatterers can be considered to be turned off and hence 
irrelevant to our self-interference mechanism. For this reason, one can
effectively realize the position tuning of the scatterers by adjusting their
resonance frequencies. % Thus, if there are
%two groups of scatterers at two different positions in the waveguide (one's $%
%\Delta x$ is odd and the other's $\Delta x$ is even), the controllable
%switching between superradiance and subradiance can be achieved by tuning
%the frequencies of these two groups of scatterers.

Finally, we discuss how the main idea of this work can be realized on an
experimental platform consisting of superconducting transmission line
resonators and transmon qubits. The coplanar transmission line resonators,
which offer the continuum for the coherent transport of nonlinear photons,
can be constructed by many equal transmission-line segments coupling with
each other by dielectric materials \cite{Zhou08,ZhouPRA08}. The transmon
qubit is viewed as a Cooper pair box shunted by a capacitor that is large
relative to the stray capacitance of the Josephson junction \cite{Koch07}.
The frequencies of transmon qubits can be manipulated via the external
magnetic flux intersecting the loop formed by the SQUID \cite%
{Koch07,Sillanpaa07}. The practicable values for the coupling
strength between resonator and transmon qubits range from a few to hundreds
of MHz while the qubit frequency can be controlled from a few to tens of GHz
that is similar to the range of the resonance frequency of each resonator 
\cite{Wendin17,Blais21}.

%\begin{acknowledgments}
\emph{Acknowledgments.}--- J.G. acknowledges fund support by the Singapore
Ministry of Education Academic Research Fund Tier-3 grant No.
MOE2017-T3-1-001. % and by the Singapore NRF Grant No. NRFNRFI2017-04. 
%\end{acknowledgments}

\onecolumngrid
\clearpage
%\newpage
%\widetext

%\begin{center}
%\textbf{\large Supplemental Material}
%\end{center}
%
%%%%%%%%%%% Merge with supplemental materials %%%%%%%%%%
%%%%%%%%%%% Prefix a "S" to all equations, figures, tables and reset the counter %%%%%%%%%%
%\setcounter{equation}{0} \setcounter{figure}{0} %\setcounter{table}{0}
%%\setcounter{page}{1}
%\makeatletter
%
%%\renewcommand{\thefigure}{SM\arabic{figure}} \renewcommand{\thesection}{SM%
%%\arabic{section}} \renewcommand{\theequation}{SM\arabic{equation}}
%\newpage
%\widetext

\begin{center}
\textbf{\large Supplementary material for: \\[0pt]
Coherent Control of Collective Spontaneous Emission through self-interference }
\end{center}

%%%%%%%%%% Merge with supplemental materials %%%%%%%%%%
%%%%%%%%%% Prefix a "S" to all equations, figures, tables and reset the counter %%%%%%%%%%
\setcounter{equation}{0} \setcounter{figure}{0} %\setcounter{table}{0}
\makeatletter

\renewcommand{\thefigure}{SM\arabic{figure}} \renewcommand{\thesection}{SM%
\arabic{section}} \renewcommand{\theequation}{SM\arabic{equation}}

\section{Dynamics of collective Spontaneous emission}

The time evolution of the amplitudes defined in Eq. (\ref{OriDynFunc}) of
the main text is given by%
\begin{equation}
	i\frac{\partial }{\partial t}C_{j}^{i}\left( t\right) =\Omega
	_{j,i}C_{j}^{i}\left( t\right) +\sum_{k}\frac{V_{j,i}e^{ikx_{\left( i\right)
	}}}{\sqrt{N}}C_{k}\left( t\right) \text{,}  \label{Dynamics1}
\end{equation}%
\begin{equation}
	i\frac{\partial }{\partial t}C_{k}\left( t\right) =\omega _{k}C_{k}\left(
	t\right) +\sum_{j,i}\frac{V_{j,i}e^{-ikx_{\left( i\right) }}}{\sqrt{N}}%
	C_{j}^{i}\left( t\right) \text{,}  \label{Dynamics2}
\end{equation}%
where $x_{\left( A\right) }=x_{1}$ and $x_{\left( B\right) }=x_{2}$. To
proceed further one may invoke the Wigner-Weisskopf and Markovian theories
by neglecting the contributions of bound states, but this treatment would
not be able to capture the fractional trapping \cite{SJohn94,SQiao19}. To go
beyond these approximations, we take the Laplace transform of Eqs. (\ref%
{Dynamics1}) and (\ref{Dynamics2}), yielding 
\begin{equation}
	i\left[ -C_{j}^{i}\left( 0\right) +s\tilde{C}_{j}^{i}\left( s\right) \right]
	=\Omega _{j,i}\tilde{C}_{j}^{i}\left( s\right) +\sum_{k}\frac{%
		V_{j,i}e^{ikx_{\left( i\right) }}}{\sqrt{N}}\tilde{C}_{k}\left( s\right) 
	\text{,}  \label{LDeq1}
\end{equation}%
\begin{equation}
	i\left[ -C_{k}\left( 0\right) +s\tilde{C}_{k}\left( s\right) \right] =\omega
	_{k}\tilde{C}_{k}\left( s\right) +\sum_{j,i}\frac{V_{j,i}e^{-ikx_{\left(
				i\right) }}}{\sqrt{N}}\tilde{C}_{j}^{i}\left( s\right) \text{.}
	\label{LDeq2}
\end{equation}%
The expression of $\tilde{C}_{j}^{i}\left( s\right) $ is written as%
\begin{equation}
	i\left[ -C_{j}^{A}\left( 0\right) +s\tilde{C}_{j}^{A}\left( s\right) \right]
	=\Omega _{A}\tilde{C}_{j}^{A}\left( s\right) +\frac{V_{A}}{N}\sum_{k}\frac{1%
	}{is-\omega _{k}}\left[ \sum_{j^{\prime }}V_{A}\tilde{C}_{j^{\prime
	}}^{A}\left( s\right) +\sum_{j^{\prime }}V_{B}e^{-ik\Delta x}\tilde{C}%
	_{j^{\prime }}^{B}\left( s\right) \right] \text{,}  \label{eq1a}
\end{equation}%
\begin{equation}
	is\tilde{C}_{j}^{B}\left( s\right) =\Omega _{B}\tilde{C}_{j}^{B}\left(
	s\right) +\frac{V_{B}}{N}\sum_{k}\frac{1}{is-\omega _{k}}\left[
	\sum_{j^{\prime }}V_{A}e^{ik\Delta x}\tilde{C}_{j^{\prime }}^{A}\left(
	s\right) +\sum_{j^{\prime }}V_{B}\tilde{C}_{j^{\prime }}^{B}\left( s\right) %
	\right] \text{.}  \label{eq1b}
\end{equation}%
Substituting $\tilde{C}_{j}^{B}\left( s\right) $ in Eq. (\ref{eq1a}) into
Eq. (\ref{eq1b}) and denoting the initial excited QEs by $j_{n}$, so $\tilde{%
	C}_{j_{n}}^{A}\left( s\right) $ can be found as the following:%
\begin{align}
	\tilde{C}_{j_{1}}^{A}\left( s\right) & =i\frac{K_{A}\left( s,M_{A}-1\right)
		K_{B}\left( s,M_{B}\right) -\left( M_{A}-1\right) M_{B}\left[
		V_{A}V_{B}F\left( s,\Delta x\right) \right] ^{2}}{\left( is-\Omega
		_{A}\right) \left\{ K_{A}\left( s,M_{A}\right) K_{B}\left( s,M_{B}\right)
		-M_{A}M_{B}\left[ V_{A}V_{B}F\left( s,\Delta x\right) \right] ^{2}\right\} }%
	C_{j_{1}}^{A}\left( 0\right)  \notag \\
	& +i\frac{K_{B}\left( s,M_{B}\right) V_{A}^{2}F\left( s,0\right) +M_{B}\left[
		V_{A}V_{B}F\left( s,\Delta x\right) \right] ^{2}}{\left( is-\Omega
		_{A}\right) \left\{ K_{A}\left( s,M_{A}\right) K_{B}\left( s,M_{B}\right)
		-M_{A}M_{B}\left[ V_{A}V_{B}F\left( s,\Delta x\right) \right] ^{2}\right\} }%
	C_{j_{2}}^{A}\left( 0\right) \text{,}  \label{Laplace}
\end{align}%
where $K_{i}(s,M_{i})=is-\Omega _{i}-M_{i}V_{i}^{2}F(s,0)$. $F(s,x)$ are
given by%
\begin{equation}
	F\left( s,x\right) =\frac{\left( is-is\sqrt{\left( s^{2}+4J^{2}\right) /s^{2}%
		}\right) ^{\left\vert x\right\vert }}{is\sqrt{\left( s^{2}+4J^{2}\right)
			/s^{2}}\left( 2J\right) ^{\left\vert x\right\vert }}\text{.}
\end{equation}%
In the calculations above, we have used the formula \cite{SEconomou79} 
\begin{equation*}
	\frac{1}{2\pi }\int dk\frac{e^{ikx}}{z+2J\cos \left( k\right) }=\frac{\left(
		-\frac{z}{2J}+\frac{z}{2J}\sqrt{1-\left( \frac{2J}{z}\right) ^{2}}\right)
		^{\left\vert x\right\vert }}{z\sqrt{1-\left( \frac{2J}{z}\right) ^{2}}}\text{%
		.}
\end{equation*}%
Without loss of generality, we set $\omega _{c}$ to be zero as reference
energy. By exchanging the positions of $C_{j_{1}}^{A}(0)$ and $%
C_{j_{2}}^{A}(0)$ in Eq. (\ref{Laplace}), one can analogously obtain the
time evolution expression of $\tilde{C}_{j_{2}}^{A}(s)$. The amplitude $%
C_{j_{n}}^{A}(t)$ is given by the inverse Laplace transform $%
C_{j}^{i}(t)=(1/2\pi i)\int_{\sigma -i\infty }^{\sigma +i\infty }\tilde{C}%
_{j}^{i}(s)e^{st}ds$. To evaluate this integral, we consider the analytic
behavior of $\tilde{C}_{j_{1}}^{A}(s)$ in the whole complex plane except a
branch cut along the imaginary axis from $-i2J$ to $i2J$. With the residue
theorem, the time dependence of $C_{j_{1}}^{A}\left( t\right) $ can then be
obtained: 
\begin{align}
	C_{j_{1}}^{A}\left( t\right) & =\frac{L_{1}(s)C_{j_{1}}^{A}\left( 0\right)
		+L_{2}(s)C_{j_{2}}^{A}\left( 0\right) }{G(s)}e^{st}|_{s=-i\Omega
		_{A}}+\sum_{m}\frac{iL_{2}(s)\left[ C_{j_{1}}^{A}\left( 0\right)
		+C_{j_{2}}^{A}\left( 0\right) \right] }{\left( is-\Omega _{A}\right)
		[G(s)]^{\prime }}e^{st}|_{s=x_{m}^{\left( 2\right) }}  \notag \\
	& -\sum_{\alpha =\pm }\int_{-1}^{1}\frac{\left[ Q_{1}^{\alpha
		}C_{j_{1}}^{A}\left( 0\right) +Q_{2}^{\alpha }C_{j_{2}}^{A}\left( 0\right) %
		\right] e^{i2Jyt}}{2\pi i\alpha \left( y+\Omega _{A}\right) [Q_{1}^{\alpha
		}-Q_{2}^{\alpha }]}dy\text{,}  \label{SDynamics}
\end{align}%
where $L_{1}(s)$ and $L_{2}(s)$ are defined as 
\begin{equation*}
	L_{1}(s)=K_{A}(s,M_{A}-1)K_{B}(s,M_{B})-(M_{A}-1)M_{B}[V_{A}V_{B}F(s,\Delta
	x)]^{2}
\end{equation*}%
and 
\begin{equation*}
	L_{2}(s)=K_{B}(s,M_{B})V_{A}^{2}F(s,0)+M_{B}[V_{A}V_{B}F(s,\Delta x)]^{2}%
	\text{.}
\end{equation*}%
The function $G(s)=L_{1}(s)-L_{2}(s)$ and $[G(s)]^{\prime }$ means the
derivative of $G(s)$ with respect to $s$. Here $x_{m}$ is pure imaginary and
the equation $G(-iE)=0$ is nothing but the system's eigenenergy equation of
the localized photon-QE dressed states. The functions $Q_{1}^{\pm }$ and $%
Q_{2}^{\pm }$ are given by 
\begin{equation*}
	Q_{1}^{\pm }=U_{A}^{\pm }(y,M_{A}-1)U_{B}^{\pm
	}(y,M_{B})-(M_{A}-1)M_{B}[V_{A}V_{B}f_{\pm }(y,\Delta x)]^{2}
\end{equation*}%
and 
\begin{equation*}
	Q_{2}^{\pm }=U_{B}^{\pm }(y,M_{B})V_{A}^{2}f_{\pm
	}(y,0)+M_{B}[V_{A}V_{B}f_{\pm }(y,\Delta x)]^{2}\text{,}
\end{equation*}%
where $U_{i}^{\pm }(y,M_{i})=-2Jy-\Omega _{i}-M_{i}V_{i}^{2}f_{\pm }(y,0)$
and $f_{\pm }(y,x)$ are given by%
\begin{equation}
	f_{\pm }(y,x)=\pm i\frac{\left( -y\pm i\sqrt{1-y^{2}}\right) ^{\left\vert
			x\right\vert }}{2J\sqrt{1-y^{2}}}\text{.}
\end{equation}%
By exchanging the positions of $C_{j_{1}}^{A}(0)$ and $C_{j_{2}}^{A}(0)$ in
Eq. (\ref{SDynamics}), one can get the time evolution of $C_{j_{2}}^{A}(t)$.

\begin{figure}[t]
	\begin{center}
		\includegraphics[width=\columnwidth]{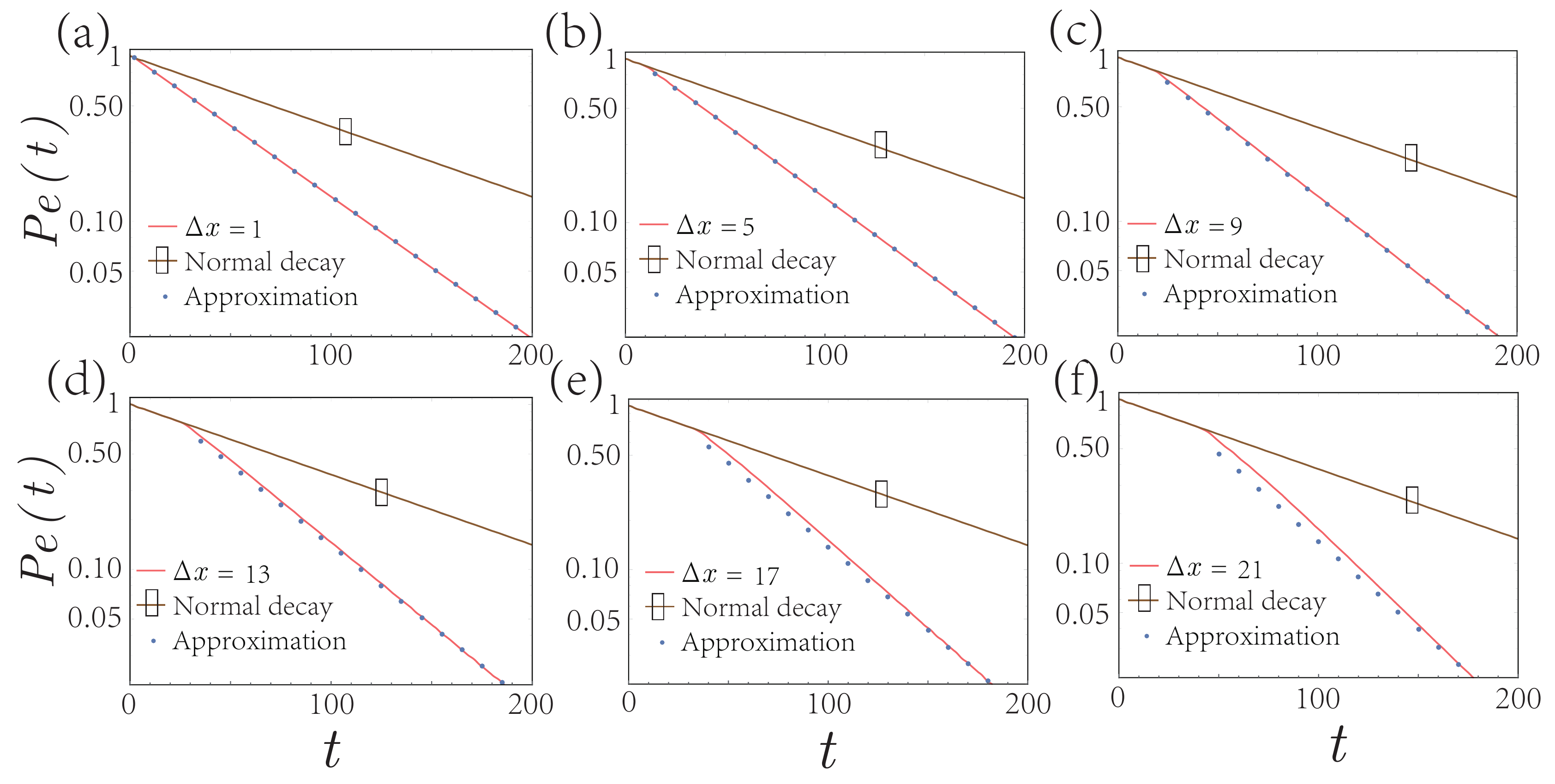}
	\end{center}
	\caption{Excited state population $P_{e}(t)=|C_{j_{1}}^{A}\left( t\right) |^{2}$ with 
		$\Delta x=1$ in (a), $\Delta x=5$ in (b), $\Delta x=9$ in (c), $\Delta x=13$
		in (d), $\Delta x=17$ in (e), $\Delta x=21$ in (f). The parameters are $%
		V_{A}/(2J)=0.07$, $V_{B}/(2J)=1.8$, $\Delta _{A}/(2J)=0$, $\Delta _{B}/(2J)=0
		$, $M_{A}=1$ and $M_{B}=2$. Time is in units of $1/(2J)$.}
	\label{fig4_deltax}
\end{figure}

For $M_{A}=1$ and $V_{A}\ll 2J$ with $\Delta x=2n+1$ ($n$ being in integer),
the first two terms in Eq. (\ref{SDynamics}) are found to be numerically much
smaller than the third term when both type-A and type-B QEs are near
resonance with a single cavity. Dropping then first two terms, $%
C_{j_{1}}^{A}\left( t\right) $ can thus  be approximately obtained as the
following: 
\begin{equation}
	C_{j_{1}}^{A}\left( t\right) \approx \frac{J}{\pi i}\int_{-1}^{1}\left\{ 
	\frac{2Jy-i\frac{M_{B}V_{B}^{2}}{2J\sqrt{1-y^{2}}}}{\left( 2Jy-i\frac{%
			V_{A}^{2}}{2J\sqrt{1-y^{2}}}\right) \left( 2Jy-i\frac{M_{B}V_{B}^{2}}{2J%
			\sqrt{1-y^{2}}}\right) +M_{B}V_{A}^{2}V_{B}^{2}\frac{\left( -y-i\sqrt{1-y^{2}%
			}\right) ^{2\Delta x}}{\left( 2J\sqrt{1-y^{2}}\right) ^{2}}}-h.c\right\}
	e^{i2Jyt}dy\text{.}
\end{equation}%
After the radiation field comes back to the decaying QE and under the
condition $\sqrt{M_{B}}V_{B}<$ or $\sim 2J$, the oscillating term $e^{i2Jyt}$
in the above expression indicates that the above integrand is insensitive to
the lower or upper limit of this integration with small $\Delta x$ (see the
following comparison with different values of $\Delta x$ between the
approximate result and exact result in Fig.~\ref{fig4_deltax}). As such,
the integration range can be safely extended to infinity. Then $%
C_{j_{1}}^{A}\left( t\right) $ can be written as 
\begin{equation}
	C_{j_{1}}^{A}\left( t\right) \approx \frac{J}{\pi i}\int_{-\infty }^{\infty
	}\left\{ \frac{2Jy-i\frac{M_{B}V_{B}^{2}}{2J}}{\left( 2Jy-i\frac{V_{A}^{2}}{%
			2J}\right) \left( 2Jy-i\frac{M_{B}V_{B}^{2}}{2J}\right)
		+M_{B}V_{A}^{2}V_{B}^{2}i^{2\Delta x}\left( 1-2\Delta xyi\right) }%
	-h.c\right\} e^{i2Jyt}dy\text{.}  \label{SCj1}
\end{equation}%
The poles of the first term in the integrand are%
\begin{align}
	y_{0,\pm }& =i\frac{1}{2}\left( \frac{M_{B}V_{B}^{2}}{4J^{2}}+\frac{V_{A}^{2}%
	}{4J^{2}}-2\Delta xM_{B}\frac{V_{A}^{2}}{4J^{2}}\frac{V_{B}^{2}}{4J^{2}}%
	\right)   \notag \\
	& \pm \frac{1}{2}\sqrt{-\left( \frac{M_{B}V_{B}^{2}}{4J^{2}}+\frac{V_{A}^{2}%
		}{4J^{2}}-2\Delta xM_{B}\frac{V_{A}^{2}}{4J^{2}}\frac{V_{B}^{2}}{4J^{2}}%
		\right) ^{2}+8M_{B}\frac{V_{A}^{2}}{4J^{2}}\frac{V_{B}^{2}}{4J^{2}}}\text{.}
\end{align}

For $\sqrt{M_{B}}V_{B}$ $\sim 2J$, $y_{0,\pm }$ are pure imaginary. Using
the residue theorem, one reduces $C_{j_{1}}^{A}\left( t\right) $ to the
following: 
\begin{equation}
	C_{j_{1}}^{A}\left( t\right) \approx \frac{M_{B}V_{B}^{2}/(4J^{2})-\alpha
		_{1}/2}{\sqrt{\beta _{1}}}e^{-J\alpha _{1}t}\text{,}  \label{SDecay1}
\end{equation}%
where $\alpha _{1}=V_{A}^{2}/[J^{2}-\Delta xV_{A}^{2}/2]$ and $\beta
_{1}=\{[2J^{2}M_{B}V_{B}^{2}-\Delta
xM_{B}V_{A}^{2}V_{B}^{2}]/(8J^{4})\}^{2}-M_{B}V_{A}^{2}V_{B}^{2}/(2J^{4})$.

In Fig.~\ref{fig4_deltax}, we compare the approximate results (dotted
lines) in Eq. (\ref{SDecay1}) with the exact results (solid lines) in Eq. (%
\ref{SDynamics}) for different values of $\Delta x$. As $\Delta x$
decreases, the approximate results agree better with the exact results.

Similar to our treatments in getting the result in Eq. (\ref{SDecay1}), in
the case of two QEs with the initial state $|\psi ^{+}\rangle =$ $(1/\sqrt{2}%
)(\sigma _{j_{1},A}^{+}+\sigma _{j_{2},A}^{+})|g$,vac$\rangle $, the
dynamics of hyperradiance for $V_{A}\ll 2J$ and $\sqrt{M_{B}}V_{B}\sim $ $2J$
can also be found with small $\Delta x$, namely%
\begin{equation}
	C_{\text{sup}}\left( t\right) \approx \frac{M_{B}V_{B}^{2}\left(
		J^{2}-\Delta xV_{A}^{2}\right) -4J^{2}V_{A}^{2}}{\left( J^{2}-\Delta
		xV_{A}^{2}\right) \sqrt{2\left( \chi ^{2}-16M_{B}V_{A}^{2}V_{B}^{2}\right) }}%
	e^{-\frac{\Gamma _{h}}{2}t}\text{,}
\end{equation}%
where the emission rate $\Gamma _{h}=4JV_{A}^{2}/(J^{2}-\Delta xV_{A}^{2})$
and $\chi =M_{B}V_{B}^{2}-\Delta xM_{B}V_{A}^{2}V_{B}^{2}/J^{2}$.

\section{Theory of photon scattering in a waveguide }

To describe the radiation field propagating towards and interacting with the quantum scatterers, we start with the Green's
function defined in terms of the system Hamiltonian $H_{\text{sca}}$ (assuming there are only scatterers but no emitters)%
\begin{equation}
G\left( z\right) =\frac{1}{z-H_{\text{sca}}}
\end{equation}%
where $z$ is a complex energy variable and the expression of $H_{\text{sca}}$ is%
\begin{equation}
H_{\text{sca}}=H_{\text{ph}}^{0}+V_{\text{ph}}
\end{equation}%
and%
\begin{equation}
H_{\text{ph}}^{0}=\sum_{k}\omega _{k}a_{k}^{\dag }a_{k}+\sum_{j}\Omega
_{B}\left\vert e_{j}^{B}\right\rangle \left\langle e_{j}^{B}\right\vert 
\text{,}
\end{equation}%
\begin{equation}
V_{\text{ph}}=\sum_{j}\sum_{k}\frac{V_{B}}{\sqrt{N}}\left( \sigma
_{j}^{B+}a_{k}e^{ikx_{2}}+\sigma _{j}^{B-}a_{k}^{\dag }e^{-ikx_{2}}\right) 
\text{.}
\end{equation}

The Green's function $G(z)$ is an analytic function on the complex $z$ plane except
at those points and branch cut of the real axis (which correspond to the
eigenvalues of bound states and scattering states). $G(z)$ satisfies the
resolvent equations \cite{SEconomou79}%
\begin{equation}
G\left( z\right) =G_{0}\left( z\right) +G_{0}\left( z\right) V_{\text{ph}%
}G\left( z\right) \text{.}
\end{equation}%
Here $G_{0}(z)$ is the free Green's function and is written as 
\begin{equation}
G_{0}\left( z\right) =\frac{1}{z-H_{\text{ph}}^{0}}\text{.}
\end{equation}%
The S-matrix element in this scattering process is given by%
\begin{equation}
S_{p,k}=\delta _{p,k}-i2\pi \delta \left( \omega _{p}-\omega _{k}\right)
T_{p,k}\left( \omega _{k}+i\epsilon \right) \text{,}
\end{equation}%
where $T(z)=V_{\text{ph}}+V_{\text{ph}}G(z)V_{\text{ph}}$ is the
T-matrix and $T_{p,k}(\omega _{k}+i\epsilon )=\left\langle g\text{,vac}%
\right\vert a_{p}T(\omega _{k}+i\epsilon )a_{k}^{\dag }\left\vert g\text{,vac%
}\right\rangle $. With the method of self-consistent equations \cite{SQiao19}%
, $T_{p,k}(\omega _{k}+i\epsilon )$ is found to be%
\begin{equation}
T_{p,k}\left( \omega _{k}+i\epsilon \right) =\frac{\frac{M_{B}V_{B}^{2}}{N}}{%
\omega _{k}-\Omega _{B}-\frac{M_{B}V_{B}^{2}}{N}\sum_{k^{\prime }}\frac{1}{%
\omega _{k}-\omega _{k^{\prime }}+i\epsilon }}\text{.}
\end{equation}%
So the S-matrix element is 
\begin{equation}
S_{p,k}=\left( 1+r_{k}\right) \delta _{p,k}+r_{k}\delta _{-p,k}
\end{equation}%
with the reflection amplitude%
\begin{equation}
r_{k}=\frac{-iM_{B}V_{B}^{2}}{\left\vert \partial \omega _{k}/\partial
k\right\vert (\omega _{k}-\Omega _{B})+iM_{B}V_{B}^{2}}\text{.}
\label{reflection}
\end{equation}%
For the case of $M_{B}=1$, Eq. (\ref{reflection}) comes back to the result
obtained in \cite{SZhou08}. The transmission amplitude can be obtained from relations $t_{k}=1+r_{k}$ and $\left\vert
r_{k}\right\vert ^{2}+\left\vert t_{k}\right\vert ^{2}=1$.

\end{document}